\documentclass[conference,a4paper]{IEEEtran}
\ifCLASSINFOpdf
   \usepackage[pdftex]{graphicx}
   \graphicspath{{Images/}}
\else
\fi
\usepackage{amssymb}
\usepackage[cmex10]{amsmath}

\usepackage{mdwmath}
\usepackage{mdwtab}

\usepackage[tight,footnotesize]{subfigure}
\usepackage[nolist]{acronym}

\begin{document}

\begin{acronym}
\acro{A/A}{Air-to-air}%
\acro{A-EXEC}{Auto Execute COCR Service}%
\acro{AAC}{Airline Administrative Control}%
\acro{ACK}{Acknowledgement}%
\acro{ACP}{Aeronautical Communications Panel}%
\acro{AF}{Assured Forwarding}%
\acro{AL}{Application Layer}%
\acro{ANSP}{Air Navigation Service Provider}%
\acro{AOC}{Airline Operational Control}%
\acro{APC}{Air Passenger Communication}%
\acro{APT}{Airport}%
\acro{ATC}{Air Traffic Control}%
\acro{ATI}{Access Technology Independent}%
\acro{ATM}{Air Traffic Management}%
\acro{ATN}{Aeronautical Telecommunication Network}%
\acro{ATN/IPS}{Aeronautical Telecommunication Network / Internet Protocol Suite}%
\acro{ATN/OSI}{Aeronautical Telecommunication Network / Open Systems Interconnection Reference Model}%
\acro{ATS}{Air Traffic Services}%
\acro{ATSC}{Air Traffic Service Communication}%
\acro{AVBDC}{Absolute Volume Based Dynamic Capacity}%
\acro{AWGN}{Additive White Gaussian Noise}%
\acro{BB}{Bandwidth Broking} %
\acro{BBFRAME}{Baseband frame}%
\acro{BE}{Best Effort}%
\acro{BER}{Bit Error Rate}%
\acro{BGAN}{Broadband Global Area Network}%
\acro{BSM}{Broadband Satellite Multimedia}%
\acro{CA}{Constrained Architecture}%
\acro{CAC}{Connection Admission Control}%
\acro{CC}{Congestion Control}%
\acro{CDMA}{Code Divison Multiple Access}%
\acro{CLNP}{Connectionless Networking Protocol}%
\acro{CLTP}{Connectionless Transport Protocol}%
\acro{CoA}{Care of Address}%
\acro{CoS}{Class of Service}%
\acro{COCR}{Communications Operating Concept and Requirements for the Future Radio System}%
\acro{COTP}{Connection Oriented Transport Protocol}%
\acro{COTS}{Commercial-Off-The-Shelf}%
\acro{CRA-CC}{CRA-Convolutional Code}%
\acro{CRA-SH}{CRA-Shannon Bound}%
\acro{CRA}{Contention Resolution ALOHA}%
\acro{CRC}{Cyclic Redundancy Check}%
\acro{CRDSA}{Contention Resolution Diversity Slotted ALOHA}%
\acro{CRDSA++}{Contention Resolution Diversity Slotted ALOHA++}%
\acro{CS}{Class Selector}%
\acro{DAMA}{Demand Assigned Multiple Access}%
\acro{DiffServ}{Differentiated Services}%
\acro{DoD}{Department of Defence}%
\acro{DSA}{Diversity Slotted ALOHA}%
\acro{DSCP}{DiffServ Code Point}%
\acro{DVB}{Digital Video Broadcasting}%
\acro{DVB-RCS+M}{Digital Video Broadcasting - Return Channel via Satellite Mobility}%
\acro{DVB-RCS}{Digital Video Broadcasting - Return Channel via Satellite}%
\acro{DVB-S2}{Digital Video Broadcasting - Second Generation}%
\acro{DVB-S2/RCS}{Digital Video Broadcasting - Second Generation / Return Channel via Satellite}%
\acro{ECN}{Explicite congestion notification}%
\acro{ECRA}{Enhanced Contention Resolution ALOHA}%
\acro{ECTL}{Eurocontrol}%
\acro{EF}{Expedited Forwarding}%
\acro{ENR}{En-Route}%
\acro{ES}{End System}%
\acro{ESA}{European Space Agency}%
\acro{ETSI}{European Telecommunications Standards Institute}%
\acro{ETSI-BSM}{European Telecommunications Standards Institute - Braodband Satellite Multimedia}%
\acro{FAA}{Federal Aviation Administration}%
\acro{FCA}{Free Capacity Assignment}%
\acro{FDM}{Frequency Division Multiplexing}%
\acro{FEC}{Forward Error Correction}%
\acro{FET}{First Entry Times}%
\acro{FL}{Forward Link}%
\acro{GEO}{Geostationary Orbit}%
\acro{GIG}{Global Information Grid}%
\acro{GSE}{Generic Stream Encapsulation}%
\acro{HA}{Home Agent}%
\acro{HF}{High Frequency}%
\acro{IC}{Interference Cancellation}%
\acro{ICAO}{International Civil Aviation Organization}%
\acro{IETF}{Internet Engineering Task Force}%
\acro{IntServ}{Integrated Services}%
\acro{IP}{Internet Protocol}%
\acro{IPS}{Internet Protocol Suite}%
\acro{IPSec}{IP Security}%
\acro{IRA}{Irregular Repetition contention resolution ALOHA}%
\acro{IRCRA}{Irregular Repetition Contention Resolution ALOHA}%
\acro{IRSA}{Irregular Repetition slotted ALOHA} %
\acro{IS}{Intermediate System}%
\acro{IS-IS-TE}{Intermediate System to Intermediate System extension for Traffic Engineering}%
\acro{ISI}{Input Stream Identifier}%
\acro{ISO/OSI}{International Organization for Standardization/Open Systems Interconnection}%
\acro{Ku}{Kurz-under}%
\acro{Ka}{Kurz-above}%
\acro{LAE}{Link Access Equipment}%
\acro{LAN}{Local Area Network}%
\acro{LD}{Low Density}%
\acro{LDACS}{L-band Digital Aeronautical Communication System}%
\acro{L-DACS}{L-band Digital Aeronautical Communication System}%
\acro{LDPC}{Low Density Parity Check Codes}%
\acro{LEO}{Low Earth Orbit}%
\acro{MAC}{Medium Access}%
\acro{MCAST}{Multicast}%
\acro{MF-TDMA}{Multifrequency Time Division Multiple Access}%
\acro{MH}{Mobile Host}%
\acro{MIHF}{Media Independent Handover Function}%
\acro{MLPP}{Multi Level Priority Preemption}%
\acro{Mode-S}{Mode-Select}%
\acro{MPE}{Multi Protocol Encapsulation}%
\acro{MPEG2-TS}{MPEG2 - Transport Stream}%
\acro{MPLS}{Multi Protocol Label Switching}%
\acro{NCC}{Network Control Centre}%
\acro{NET}{Network Connectivity}
\acro{NEWSKY}{NEtWorking the SKY for aeronautical communications}%
\acro{NOC}{Non-Operational Communication}%
\acro{NSIS}{Next Steps in Signalling}%
\acro{OC}{Operational Communication}%
\acro{ORP}{Oceanic, Remote and Polar}%
\acro{OSI}{Open Systems Interconnection}%
\acro{OSPF}{Open Shortes Path First}%
\acro{PDF}{Probability Density Function}%
\acro{PCE}{Predictive Capacity Estimation}%
\acro{PCN}{Pre-Congestion Notification}%
\acro{PDU}{Protocol Data Unit}%
\acro{PEP}{Performance Enhancing Proxy}%
\acro{PER}{Packet Error Rate}%
\acro{PID}{Packet Identifier}%
\acro{PLFRAME}{Physical Layer Frame}%
\acro{PLR}{Packet Loss Rates}%
\acro{QoS}{Quality of Service}%
\acro{QoS-PRN}{QoS Private Relay Nodes}%
\acro{QoS-RN}{QoS-Relay Nodes}%
\acro{RA}{Random Access}%
\acro{RBDC}{Rate Based Dynamic Capacity}%
\acro{RCB}{Random Coding Bound}%
\acro{RCS}{Return Channel via Satellite}%
\acro{RCS+M}{Return Channel via Satellite plus Mobility}%
\acro{RF}{Radio Frequency}%
\acro{RFC}{Request For Comments}%
\acro{RL}{Return Link}%
\acro{RM}{Resource Management}%
\acro{RN}{Relay Nodes}%
\acro{RRM}{Radio Resource Management}%
\acro{RS}{Reed Solomon}%
\acro{RSVP}{Resource Reservation Protocol}%
\acro{RTT}{Round Trip Time}%
\acro{SA}{Slotted ALOHA}%
\acro{SANDRA}{Seamless Aeronautical networking Through Integration of Data Links, Radios and Antennas}%
\acro{SB}{Shannon Bound}%
\acro{S2}{Second Generation}%
\acro{SESAR}{Single European Sky ATM Research}%
\acro{SF}{Superframe}%
\acro{SIC}{Successive Interference Cancellation}%
\acro{SLS}{Service Level Specification}%
\acro{SNDCF}{Subnetwork Dependent Convergence Function}%
\acro{SNIR}{Signal to Noise and Interference ratio}%
\acro{SNR}{Signal-to-Noise ratio}%
\acro{SWIM}{System Wide Information Management}%
\acro{SYN}{Synchronization}%
\acro{SYNC}{Synchronization}%
\acro{TCP}{Transport Control Protocol}%
\acro{TCP/IP}{Transport Control Protocol / Internet Protocol Suite}%
\acro{TDMA}{Time Division Multiple Access}%
\acro{TE}{Traffic Engineering}%
\acro{TI}{Technology Independent}%
\acro{TI-SAP}{Technology Independent - Service Access Point}%
\acro{TD}{Technology Dependent}%
\acro{TL}{Transport Layer}%
\acro{TMA}{Terminal Maneuvering Area}%
\acro{ToS}{Type of Service}%
\acro{UCA}{Unconstrained Architecture}%
\acro{UDP}{User Datagram Protocol}%
\acro{VBDC}{Volume Based Dynamic Capacity}%
\acro{VDL}{VHF Digital Mode}%
\acro{VHF}{Very High Frequency}%
\acro{VoIP}{Voice over IP}%
\acro{VPN}{Virtual Private Networks}%
\acro{WG}{Working Group}%
\acro{WiMAX}{Worldwide Interoperability for Microwave Access}%
\acro{XFECFRAME}{Forward Error Correction Frame}%
\end{acronym}

%
\title{Enhanced Contention Resolution Aloha - ECRA}

\author{\IEEEauthorblockN{Federico Clazzer and Christian Kissling}
\IEEEauthorblockA{German Aerospace Center (DLR)}
\IEEEauthorblockA{Oberpfaffenhofen, D-82234, Wessling, Germany}
\IEEEauthorblockA{Email: \{federico.clazzer,
christian.kissling\}@dlr.de} }

%


\maketitle

\begin{abstract}
\ac{RA} \ac{MAC} protocols are simple and effective when the
nature of the traffic is unpredictable and random. In the following
paper, a novel \ac{RA} protocol called \ac{ECRA} is presented. This
evolution, based on the previous \ac{CRA} protocol, exploits the
nature of the interference in unslotted Aloha-like channels for
trying to resolve most of the partial collision that can occur
there. In the paper, the idea behind \ac{ECRA} is presented together
with numerical simulations and a mathematical analysis of its
performance gain. It is shown that relevant performance increases in
both throughput and \ac{PER} can be reached by \ac{ECRA}
with respect to \ac{CRA}. A comparison with \ac{CRDSA} is also
provided.
\end{abstract}


%
\IEEEpeerreviewmaketitle


\section{Introduction}

In the recent past \ac{RA} \ac{MAC} protocols are attracting
increasing interest in many different fields, from car-to-car
communication to underwater sensor networks, just to mention a few.
RA protocols are especially suitable for all the scenarios where the
traffic is unpredictable and completely random or in cases where
only small data volumes and urgent data need to be transmitted and
\ac{DAMA} would cause delay and signalling overhead.

The current RA protocols have evolved significantly from the
original idea of Aloha proposed by Abramson in 1970
\cite{Abramson1970}. First the well known slotted evolution of Aloha
has been presented and analyzed by Roberts \cite{Roberts1975} few
years later.

In the last years \ac{CRDSA} \cite{Casini2007} and \ac{CRA}
\cite{Kissling2011a} have been presented as very promising
evolutions of respectively Slotted Aloha and Aloha. \ac{CRDSA},
is an evolution of the Slotted Aloha scheme and in
particular of \ac{DSA} proposed in \cite{Choudhury1983}.
\ac{DSA} provides a lower delay and higher throughput than \ac{SA}
under very moderate loading conditions by transmitting twice the
same packet in a different \ac{TDMA} slot, or a different frequency
and time slot in case of Multi-Frequency \ac{TDMA}. However, the
throughput difference between Aloha and Slotted Aloha or \ac{DSA}
was limited and quite poor in absolute terms.

\ac{CRDSA} takes from \ac{DSA} the idea to send more than one packet
instance per user for each frame. The original \ac{CRDSA} protocol
generates two replicas of the same packet at random times within the
frame instead of only one as in \ac{SA}. While the driver for
\ac{DSA} is to slightly enhance the \ac{SA} performance by
increasing the probability of successful reception of one of the
replicas (at the expense of increased random access load),
\ac{CRDSA} in addition is designed in a way to resolve most of
\ac{DSA} packet contentions by using \ac{SIC}. Packet collisions are
cleared up through an effective \ac{SIC} approach that uses frame
composition information from the replica packets. The key idea of
\ac{CRDSA} is to provide in each replica the signaling information
of where the other replicas of the corresponding user are placed in
the frame. Every time a packet is recovered, this information can be
used for removing the signal contribution of the other replica from
the frame, thus possibly removing its interference contribution to
other packets. The main \ac{CRDSA} advantages compared to Slotted Aloha
lie in an improved \ac{PLR} and a much higher operational throughput.

The \ac{CRA} protocol exploits the same approach of using \ac{SIC}
as \ac{CRDSA}, but in an Aloha-like \ac{MAC} protocol. Unlike the
slotted schemes, here no slots are present in the frame and thus the
replicas of the users can be placed within the frame without
constraints, except that replicas of a user may not interfere each
other. The avoidance of slots results in significant advantages such
as relaxation in synchronization requirements among users and
possibility of varying packet length without padding overhead.
\ac{FEC} in \ac{CRA} is, unlike in \ac{CRDSA}, beneficial also when
no power unbalance among users is present because partial
interference is not only possible but also more probable than
complete interference.

The \ac{IRSA} \cite{Liva2011} protocol evolution is a bipartite
graph optimization of \ac{CRDSA}, where the number of replicas for
each user is not fixed but is taken from a probability distribution
for maximizing the throughput. It was shown in \cite{Liva2011} that
the distribution can be optimized to either maximize the throughput
or to minimize the \ac{PLR}.

The performance evaluations within
\cite{RioHerrero2009}-\cite{RioHerrero2008} have shown that the
maximum throughput of \ac{CRDSA} (normalized to slots) can be
impressively extended from $T_{SA}=0.36$ (where
$T_{SA}$ is the normalized throughput of Slotted Aloha), up to
$T_{CRDSA}\cong 0.55$ and even up to
$T_{CRDSA++}\cong 0.68$ when 4 replicas per user are
sent. With the \ac{IRSA} approach a maximum theoretical throughput
of $T_{IRSA}=0.97$ can be achieved with a
distribution obtained via differential evolution \cite{Liva2011}
containing 16 replicas per user at maximum . In
\ac{CRA}, assuming \ac{FEC} $=1/2$ and QPSK modulation, the maximum
theoretical throughput shown with two replicas per user is
$T_{CRA}=0.96$ using the Shannon's capacity limit. Moreover, the
\ac{PLR} drops down to very low values.
While \ac{SA} meets a \ac{PER} of $10^{-3}$ for a normalized offered
traffic load $G=10^{-3}\,Erl$, the \ac{CRDSA} scheme can meet the
same \ac{PLR} for offered traffic load of $G=5.5\cdot 10^{-2}\,Erl$
and \ac{CRDSA}++ even for $G=0.6\,Erl$. The \ac{IRSA} protocol
attains a \ac{PLR} of $10^{-3}$ for $G=0.3\,Erl$ while the \ac{CRA}
protocol obtains the same \ac{PLR} for $G=0.6\,Erl$ with \ac{FEC}
$=1/2$ and QPSK modulation.

\section{Partial Interference and Loops}

When slotted schemes like \ac{CRDSA} are considered, for each packet
sent in the frame only two cases are possible, either no
interference between packets or entire overlapping. Moving to
unslotted schemes like CRA different interference levels among
packets are possible due to the elimination of the slots and random
starting times. In this case, an entire overlapping between packets
is only one possible interference scenario.

However, the \ac{SIC} process can get stuck in situations where the
replicas of two or several users interfere with each other in a way
that none of the replicas can be recovered and thus all involved
packets are lost. Such a case is denoted as a \textit{loop}. While
the probability of having such \textit{loops} decreases with
increasing frame length, practical frame lengths have a non
negligible probability of \textit{loops}.

\begin{figure}
\centering
\subfigure[CRDSA loop]{
    \includegraphics[width=8cm]{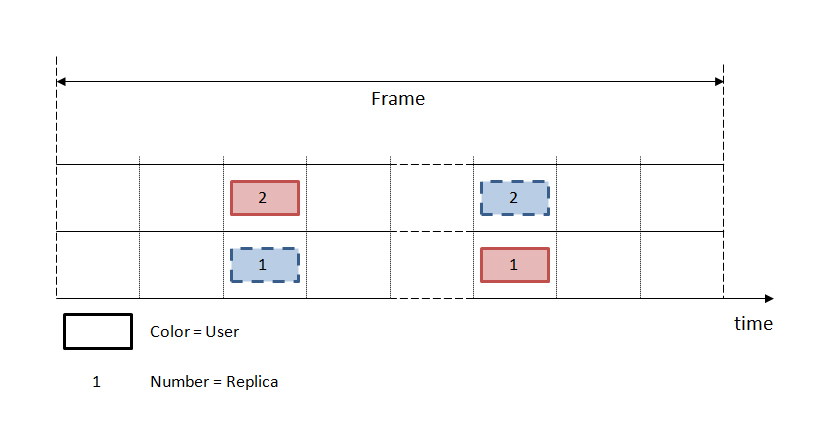}
    \label{fig:loop_a}
}\\
\subfigure[CRA loop]{
    \includegraphics[width=8cm]{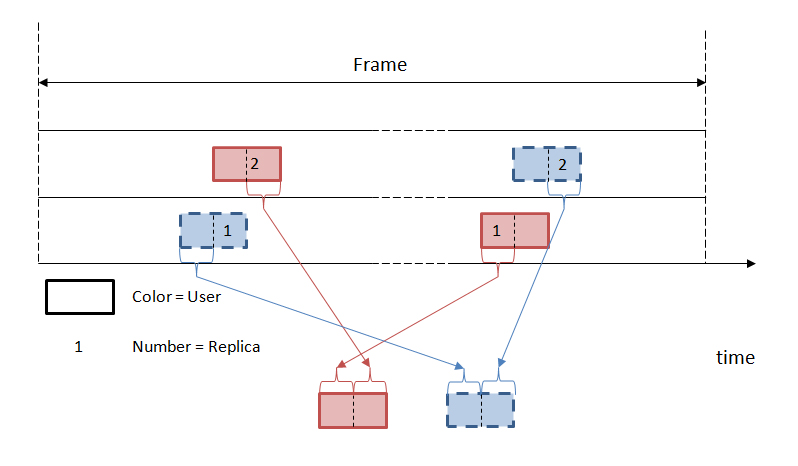}
    \label{fig:loop_b}
}
\caption{\ac{CRDSA} and \ac{CRA} simplest loops}
\label{loops}
\end{figure}

In Fig. \ref{fig:loop_a} the simplest \textit{loop} that can occur
in case of \ac{CRDSA} is presented. In the situation shown,
if the degree $d$ is equal to 2, both users cannot be decoded since
both replicas of each user are fully interfered by the replica of
the other user.

When \ac{CRA} is considered, the interference might not be
completely destructive for the users involved. In fact, since no
slots exist here, partial interference among users can occur and is
more probable than a complete interference. If the interference
experienced by a replica is sufficiently small and the error
correction code is strong enough, the packet can still be correctly
decoded. However there are wide number of cases where this is not
possible, in particular if the interference power is too high. E.g.,
if we consider the scenario presented in Fig.
\ref{fig:loop_b} and we suppose that the interference power is too
high for being corrected by the \ac{FEC} code, then in \ac{CRA} both
replicas of the user cannot be correctly decoded although different
parts of the two replicas are affected by interference.

If it would be possible to combine the uninterfered symbols of the
replicas of a users into a new packet, then the new combined packet
might be decoded successfully and unlock the loop. In this case the
two users which could not be decoded in \ac{CRA} can now be
correctly decoded and removed from the frame. In the case presented
in Fig. \ref{fig:loop_b}, if we take the uninterfered first 50\% of
the blue dashed users first replica and the uninterfered second 50\%
of the second replica of the same user and we combine them creating
a new packet, we could obtain a packet free of interference. The red
user can then also be recovered using the same procedure and
creating the combined packet. Creating a new combined packet from
the replicas may however not work in all situations. If the same
parts of all replicas of a user are interfered, the combined packet
will not have a higher \ac{SNIR} and the loop could not be resolved.

In \cite{ZigZag} a similar scenario has been addressed, but the proposed
solution exploits an iterative \textit{chunk-by-chunk} decoding between
the collided packets where decoding errors propagate, while in \ac{ECRA}
a combined packet is constructed and the decoding attempted on it in one
step. Moreover, in \ac{ECRA} the replicas generation is made regardless
of the decoding success, while in \cite{ZigZag} only the collided packets are
replicated. \cite{SigSag} can be seen as the soft-decoding version
of \cite{ZigZag}. The decoding algorithm of \cite{SigSag} propagates
the probabilities associated to the received symbols, instead of
hard-decoding the symbols and use them for the back-substitution.
Practical implementations issues arise in this second version,
e.g. the enabling of bit permutations, how to access the soft
information and finally the increase of complexity compared to
\cite{ZigZag}.

\section{ECRA decoding procedure}

\ac{ECRA} follows a two steps procedure for decoding the packets at
the receiver side. At first the current frame is stored and the
\ac{SIC} is applied on the received packets. The \ac{SIC} begins to
scan the frame from the first received symbol and once it finds a
packet it tries to decode it. If the decoding was successful, the
content of the packet payload can be recovered. Since every packet
contains information on the position of the current user replica(s),
we can exploit this information for removing the other replica(s)
from the frame.

In the following we assume
ideal interference cancellation for this. If the decoding was not
successful the packet remains in the frame and the interference
contribution is not removed. Independently from the correct or
incorrect decoding of the previous packet, the \ac{SIC} pursues to
scan the frame looking for the next packet. When the end of the
frame is reached, either all the users packets have been
correctly decoded or the replicas of at least two users are still
not decoded and thus present in the frame. Hence, in the former the
\ac{SIC} procedure stops, while in the latter the \ac{SIC} procedure
tries to scan again the frame from the beginning. The \ac{SIC}
procedure is stopped if either all packets have been successfully
decoded, if no further packets could be decoded in a round or if the
maximum number of interference cancellation rounds has been reached.

The second step is the key novelty of the presented \ac{ECRA}
protocol. For each remaining user in the frame, the replicas
sections without interference are taken and are used for creating a
new combined packet for the considered user (see also Fig.
\ref{fig:loop_b}). If some portions of the user packets encounter
interference in all the replicas, the replica symbols with the
lowest interference are taken and exploited for creating the
combined packet. This leads to create a user packet with the
lowest possible interference. On the combined packet, the decoding
is attempted and if the packet is correctly received, it is
re-encoded modulated and removed in all the positions within the
frame where the replicas were placed. Like in the first step, the
mentioned procedure is iterated until either all the users can be
correctly received or the maximum of possible iterations is reached.

It is possible to show that the \ac{ECRA} approach can always
generate a packet with higher, or at least equal, \ac{SNIR}
with respect to \ac{CRA}. Given the packet of user $u$ and replica
$r$ positioned within the frame and selected its symbol $s$, we can
compute the interference contribution $I_{s,u,r}$ of other users
replicas in symbol $s$ to the replica $r$ of user $u$:

\begin{equation}
\label{31}
I_{s,u,r}=\sum_{i=1,i\neq u}^{t_u}\sum_{d=1}^{deg}\delta_{id}, \,with \, \left\{
\begin{array}{rl}
\delta_{id} = 1 & \mbox{if replica $d$ of }\\
                & \mbox{user $i$ has a symbol}\\
                & \mbox{at position $s$ of}\\
                & \mbox{user $u$, replica $r$}\\
\delta_{id} = 0 & \mbox {otherwise}
\end{array}
\right .
\end{equation}

where $t_u$ is the number of users in the frame and $deg$ is the number of replicas per user.
The interference ratio suffered by the given packet $x_{u,r}$ is
then:

\[
x_{u,r}=\frac{1}{t_s}\sum_{s=1}^{t_s}I_{s,u,r}
\]

where $t_s$ is the number of symbols in the packet of user $u$ and
replica $r$. Under the assumption of equal power conditions among
users, the SNIR of user $u$ and replica $r$ gets:

\[
SNIR_{u,r}=\frac{P}{x_{u,r} \cdot P+N}=\frac{SNR}{x_{u,r} \cdot SNR + 1},
\]
with the transmission power $P$ and noise power $N$.

For example, if the replica $r$ of user $u$ experiences an overall
interference of 50\%, then $x_{u,r}=0.5$.

Since the \ac{SNR} of the packet is known, it is possible to
evaluate the \ac{SNIR}. Given $I_{s,u,r}$ from equation \eqref{31},
for each replica $r\in\mathcal{D}_u$ with $\mathcal{D}_u$ the set of
all the user $u$ replicas, the \ac{ECRA} protocol selects the symbol
$s$ with the lowest interference $I_{s,u,r}^*$ among all symbols at
the same location in the current user replicas in $\mathcal{D}_u$

\[
I_{s,u,r}^*=\min_{r\in\mathcal{D}}\left\{I_{s,u,r}\right\}\leq I_{s,u,r}, \quad r=1,...,deg.
\]

The interference ratio suffered by the combined packet $x_{u}^*$
created by \ac{ECRA} is then:

\[
x_{u}^*=\frac{1}{t_s}\sum_{s=1}^{t_s}I_{s,u,r}^*\leq\frac{1}{t_s}\sum_{s=1}^{t_s}I_{s,u,r}=x_{u,r}, \quad r=1,...,deg
\]

Under the assumption of equal power conditions among users, the
\ac{SNIR} of the \ac{ECRA} combined packet is:

\begin{multline}
\label{32}
SNIR_{u}^*=\frac{P}{x_{u}^* \cdot P + N}\geq\frac{P}{x_{u,r} \cdot P+N}=SNIR_{u,r},\\
r=1,...,deg.
\end{multline}

\begin{figure}
\centering
\includegraphics[width=8.5cm]{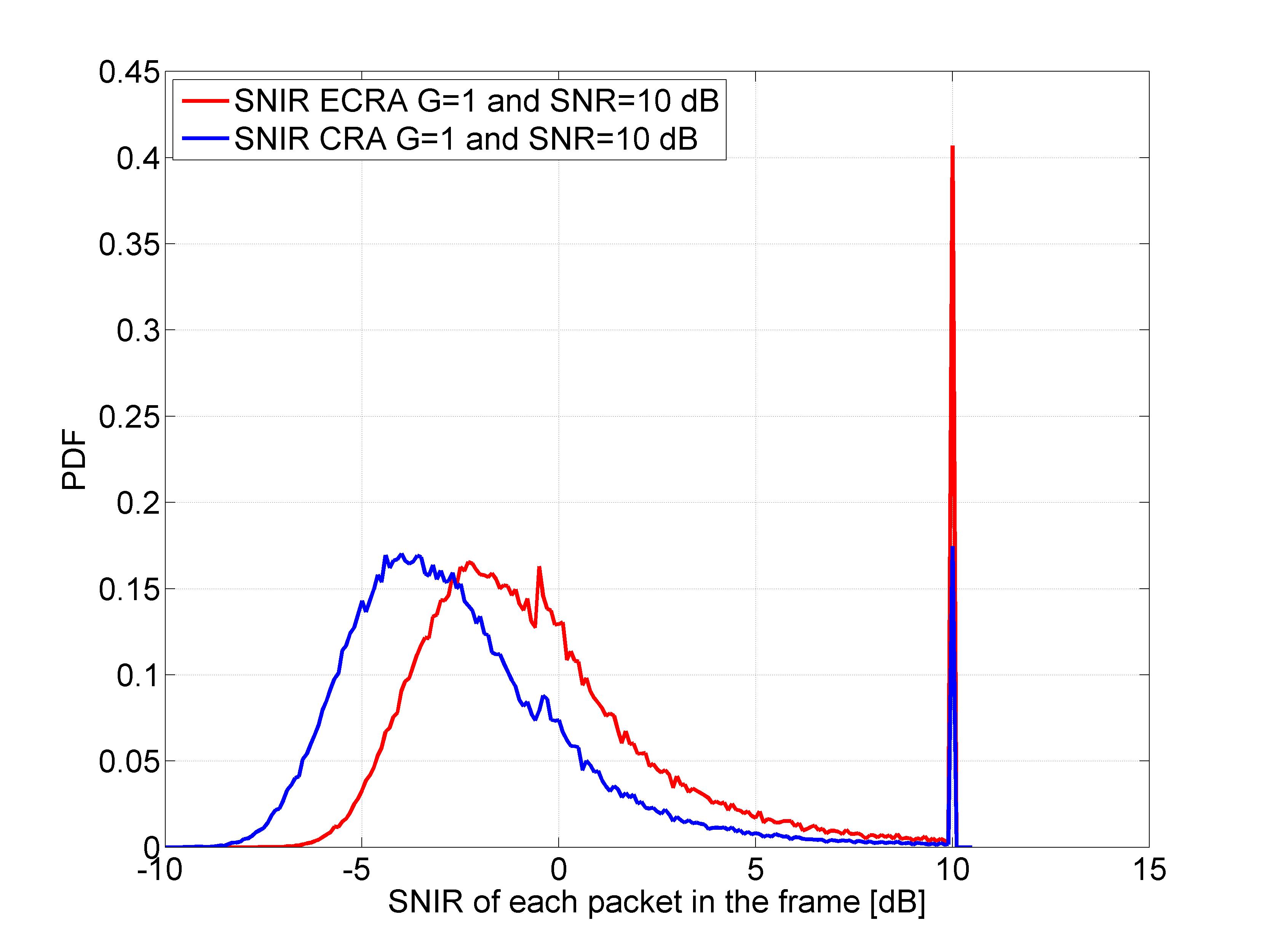}
\caption{\ac{SNIR} Probability Density Function for \ac{ECRA} and \ac{CRA}}
\label{SNIR_PDF}
\end{figure}

When each symbol with the lowest level of interference belongs to
one single packet, the SNIR of ECRA coincides with the SNIR of CRA.
In all the other cases we have $SNIR_{ECRA}>SNIR_{CRA}$. The result
of equation \eqref{32} is confirmed by the simulations summarized in
Fig. \ref{SNIR_PDF}. The \ac{PDF} of \ac{CRA} is shifted on the left
of the graph with respect to the \ac{ECRA} \ac{PDF}. In other words,
\ac{CRA} shows a higher probability of low \ac{SNIR} packets
compared to \ac{ECRA}. It can be noted that in both cases a peak
in the PDF is found at $SNIR=10$ dB, which corresponds to the packets
free of interference. In fact, an \ac{SNR} of 10 dB was selected for the
simulations.

It is important to underline that the second step of \ac{ECRA}, needs
complete knowledge of the replicas position of the remaining users in
the frame. Under this assumption, it is always possible to create the
combined packet with the lowest level of interference, because the collided
packets portions are known. For the knowledge of the frame composition in
practical systems the replica location information stored in packet headers,
which are protected with a very robust \ac{FEC}, can be
exploited. Very robust \ac{FEC} applied to the headers can allow retrieving
the information about replica locations although the packet itself is not
decodable due to collisions. Investigations on the best position
of the signaling info within the packet, on the advantage of
replicating this info and/or the use of dedicated correction codes
will be addressed in future work. It will be shown that
compared to the case of perfect frame knowledge a smart
positioning of this info within the packet results only in a
minor performance degradation.

The first step of the \ac{ECRA} protocol applies the \ac{SIC}
procedure exploited also in \ac{CRA} while the second step increases
the probability of correct decoding of the packets by generating new
combined packets, but at the cost of increasing the complexity of
the decoder. There are some situations where it may be more
important to have less complex receivers even if they have the
drawback of decreased performance, but in other scenarios it may be
necessary to exploit the maximum performance. In the latter case,
\ac{ECRA} is a superior technique compared to \ac{CRA}, when
unslotted Aloha-like random access \ac{MAC} protocols are
considered.

\section{Numerical Results}

Three different sets of simulations of \ac{ECRA} are shown in the
following section. The behavior of \ac{ECRA} is analyzed in terms of
\ac{SNR} and in terms of the rate $R=R_c \cdot log_2(M)$, with code
rate $R_c$  and modulation index $M$. The comparison between the
Shannon's capacity limit, called in the following \ac{SB}, and
\ac{RCB} \cite{Gallager1968} is provided as well. In both cases, the
co-user interference is assumed to be Gaussian distributed. It can
be shown that this assumption is not far from reality due
to phase-, time- and frequency offsets between the signals.

\subsection{Shannon Bound}

At first, the decoding threshold is approximated with the \ac{SB},
assuming a Gaussian channel. The correct decoding of a given packet in this case, is only related
with the quantity of interference plus noise that the packet
experiences due to collision in the \ac{MAC} frame and the noise
given by the \ac{AWGN} channel. Thanks to the Hartley-Shannon
Theorem, it is commonly known that in an \ac{AWGN} channel every
rate $R$ that accomplishes the relation $R\leq C/W = log_2(1+SNR)$,
where $C$ is the channel capacity and $W$ the channel bandwidth,
allows in theory an error free decoding. Thus the maximum allowable
rate is $R = C/W = log_2(1+SNR)$. Moving from the \ac{SNR} to the
\ac{SNIR} ratio, we find $R = C/W = log_2(1 + SNIR_{SHA})$, where
$SNIR_{SHA}$ is the decoding threshold we are looking for.
Elaborating the previous equation for extracting the $SNIR_{SHA}$ we
get $SNIR_{SHA} = 2^R - 1$.

In \ac{ECRA}, for each user $u \in \mathcal{U}$ where $\mathcal{U}$
is the set of users sending packets in a certain frame, the $SNIR_u^*$
for the combined packet of user $u$ is given by equation \eqref{32}.
Each packet with $SNIR_u^* \geq SNIR_{SHA}$ is considered to be
correctly decoded an its signal is removed from the frame as well
as all the replicas of the corresponding user. Otherwise, the
current packet remains in the frame.

\subsection{Random Coding Bound}

Moving from the theoretical limit given by the \ac{SB} which
is not reachable in practice, to a more realistic one,
leads to consideration of the \ac{RCB}. The \ac{RCB} represents the
upper bound on the average block error probability for codes of $n$
symbols and for a given rate $R$. Since the \ac{RCB}
considers the average error probability of a set of codes, we are
ensured that at least one code can reach this probability or less
\cite{Gallager1968}.

For the simulations, given the rate $R$, the
\ac{RCB} \ac{PER} over the \ac{SNR} curve is generated. The \ac{PER}
curve is then used as probability to correctly decode any given
packet with its corresponding \ac{SNIR}.

\subsection{Simulations}

The performed simulations show the average throughput $\overline{T}$
and the average packet error rate $\overline{PER}$ for a set of
traffic offered load values $G$.

The first set of simulations provided are done for a rate $R=2$
adopting the \ac{SB} as decoding threshold. The considered
scenario is characterized by a nominal $SNR = 10\,\text{dB}$ equal
for each user generating traffic, the frame duration is selected to
$T_f = 100\,\text{ms}$ and the symbol duration to $T_s =
1\,\mu\text{s}$. Moreover, the packet length $L_p = 1000$ bits is
equal for every user, the number of replicas sent within the frame
by any given user is $d = 2$ and the maximum number of \ac{SIC}
iterations for the three \ac{RA} schemes is $I_{max} = 10$.
The rate $R = 2$ leads to a packet length, in symbols, $L_s = L_p /
R = 1000 / 2 = 500$ symbols. For any given value of $G$,
$\overline{T}$ and $\overline{PER}$ are averaged over $N_f=1000\,
frames$.

We can suppose for example that \ac{FEC} is adopted and the
implemented encoder uses a code rate $R_c = 1/2$. In this case, the
modulation index must be $M = 16$ to result in a rate $R = 2$, which
corresponds to a 16-QAM modulation.

Therefore, the normalized traffic load $G$ is given by:

\[
G = \frac{N_u \cdot L_p \cdot T_s}{T_f \cdot R},
\]

with $N_u$ the number of users sending packets in the frame. The average throughput
$\overline{T}$ is defined as the probability of successful reception
of a packet, multiplied by the offered traffic load $G$. The average
throughput here is related to the logical throughput, i.e. user
packets, whereas the physical throughput would also consider the
number of replicas generated per packet. The average packet error
rate $\overline{PER}$, is evaluated as:

\[
\overline{PER} = \frac{P_{err}}{N_u \cdot N_f}
\]

where $P_{err}$ is the number of lost packets at the receiver side,
and $N_f$ is the number of simulated frames for the corresponding
$G$. Since the $\overline{PER}$ represents the average probability
of a packet error, $\overline{T}$ is computed in the following way:

\[
\overline{T} = (1-\overline{PER}) \cdot G.
\]

For simplicity of notation, the average $\overline{PER}$ and
$\overline{T}$ are denoted as $PER$ and $T$ in the following.

\begin{figure}
\centering
\includegraphics[width=9.5cm]{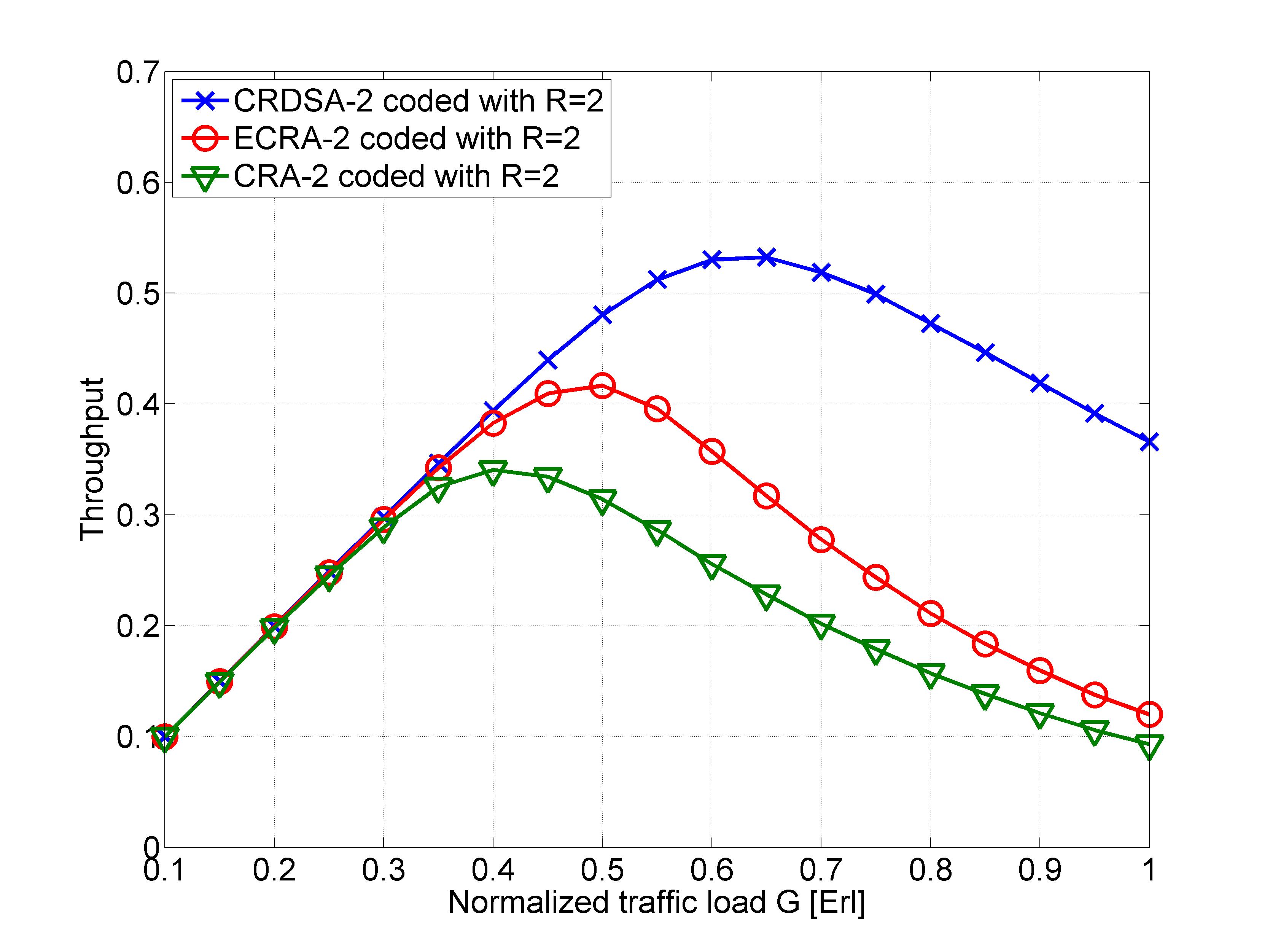}
\caption{CRDSA, CRA and ECRA throughput comparison for rate $R=2$, $SNR=10\,\text{dB}$ and \ac{SB}}
\label{ECRA_R2_T}
\end{figure}

In Fig. \ref{ECRA_R2_T} the throughput comparison of \ac{ECRA},
\ac{CRA} and \ac{CRDSA}-2 in the scenario discussed above is
presented. The maximum throughput reached by \ac{CRA} is
$T_{max-CRA}\cong 0.34$ at $G=0.4\,\text{Erl}$, while \ac{ECRA}
shows a maximum throughput of $T_{max-ECRA}\cong 0.42$ at
$G=0.5\,\text{Erl}$. The percentage of maximum throughput increase
from \ac{CRA} to \ac{ECRA} is roughly 23\%, which is a significant
improvement. Finally, \ac{CRDSA}\footnote{\ac{CRDSA}-2 is shown here
as the most basic representant of slotted \ac{SIC} schemes. It
should be noted that higher order \ac{CRDSA} can achieve better
performance than \ac{CRDSA}-2} in the same conditions is able to
reach a maximum throughput $T_{max-CRDSA}\cong 0.53$ at
$G=0.65\,\text{Erl}$. The \ac{ECRA} \ac{RA} scheme achieves a
throughput in between the one of \ac{CRA} and \ac{CRDSA} in the
region of positive slope while it shows a behavior more similar to
\ac{CRA} in the negative slope region. It is important to recall that
the better performance of \ac{CRDSA} w.r.t. to both \ac{CRA} and 
\ac{ECRA} comes at the expense of stronger synchronization
constraints at the users.

\begin{figure}
\centering
\includegraphics[width=9.5cm]{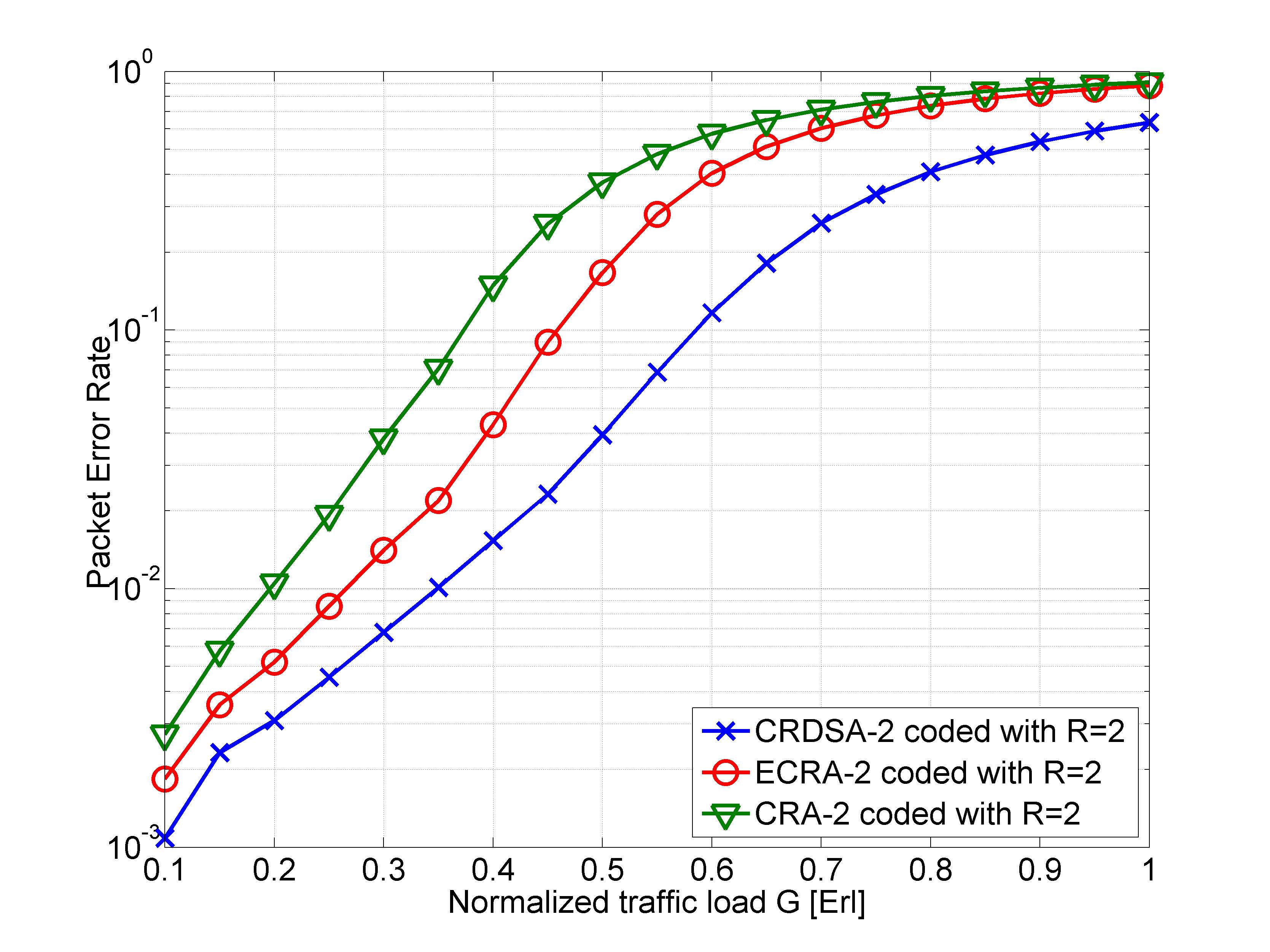}
\caption{CRDSA, CRA and ECRA packet error rate comparison for rate $R=2$, $SNR=10\,\text{dB}$ and \ac{SB}}
\label{ECRA_R2_PER}
\end{figure}

In Fig. \ref{ECRA_R2_PER} the \ac{PER} behavior of the three \ac{RA}
schemes over $G$ is presented. The colors and symbols are the
same as used in Fig. \ref{ECRA_R2_T}. For small to average values of
$G$, \ac{ECRA} \ac{PER} shows a significant improvement compared to
\ac{CRA} but it is still worse than the \ac{CRDSA} slotted scheme.
Above the value of $G = 0.55\,\text{Erl}$, the \ac{ECRA} and
\ac{CRA} \ac{PER} curves tend to converge. The minimum \ac{PER} of
the three schemes is for \ac{CRA}, $PER_{min-CRA}\cong 3 \cdot
10^{-3}$; for \ac{ECRA}, $PER_{min-ECRA}\cong 2 \cdot 10^{-3}$; and
for \ac{CRDSA}, $PER_{min-CRDSA}\cong 1 \cdot 10^{-3}$, at
$G=0.1\,\text{Erl}$ for all the schemes.

\begin{figure}
\centering
\includegraphics[width=9.5cm]{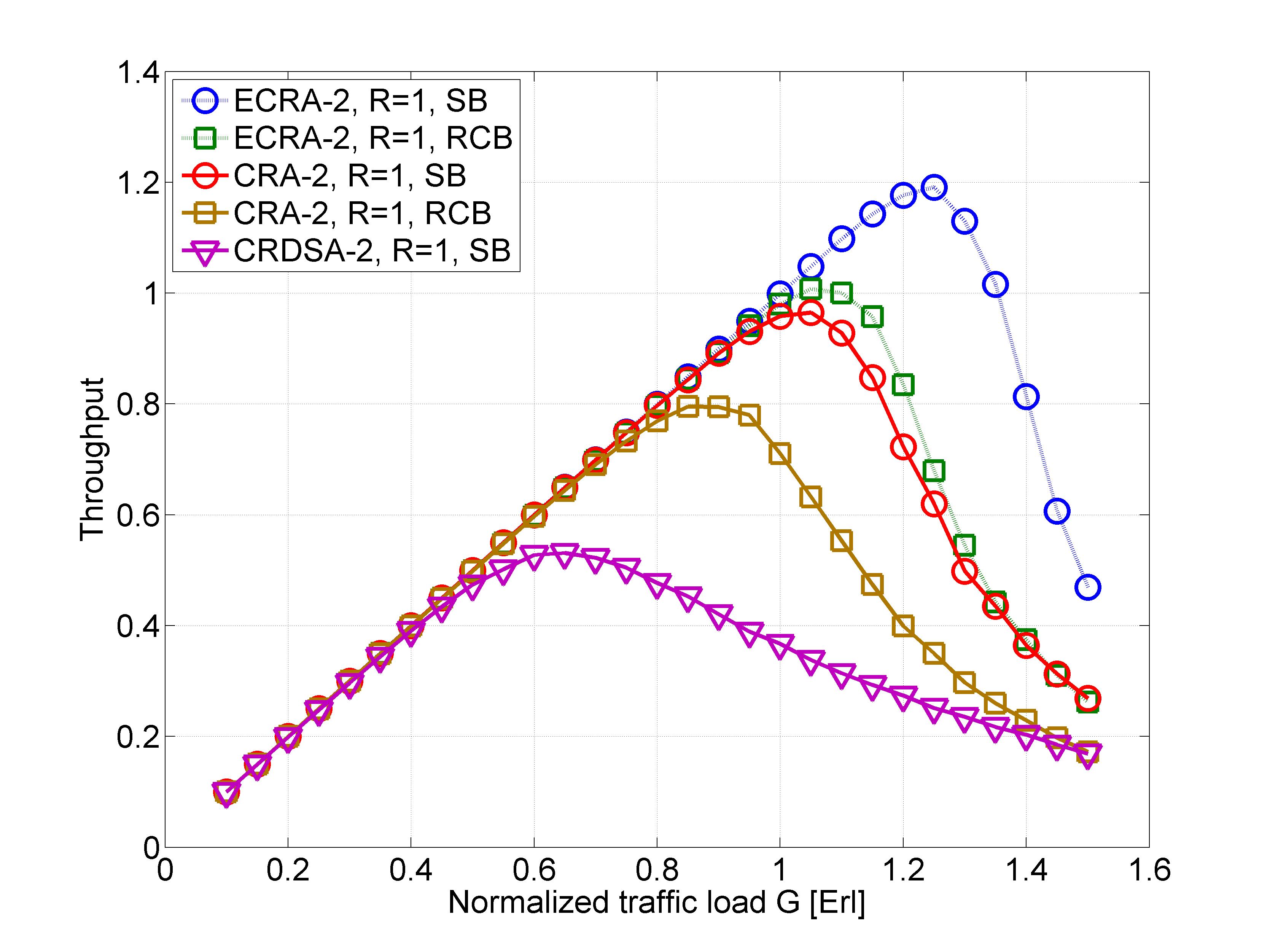}
\caption{\ac{CRDSA}, \ac{CRA} and \ac{ECRA} throughput comparison for rate $R=1$ and $SNR=10\,\text{dB}$}
\label{ECRA_R1_10_T}
\end{figure}

The second set of simulations provided are done for a rate $R=1$
comparing the \ac{SB} as decoding threshold with the \ac{RCB}. All
the other simulations parameters are the same as explained before.
In Fig. \ref{ECRA_R1_10_T} the throughput of \ac{ECRA}, \ac{CRA} and
\ac{CRDSA}\footnote{In CRDSA, $R=1$ is not the best choice from a
spectral efficiency point of view, if the $SNR = 10\,dB$ . Since we
are not interested in maximizing the spectral efficiency, the same
rate is used for all the considered schemes to have equal
conditions.} are compared. \ac{ECRA} with the \ac{SB} as decoding
threshold reaches the maximum throughput $T_{max-ECRA} = 1.19$ at $G
= 1.25\,\text{Erl}$ outperforming both \ac{CRA} and \ac{CRDSA}-2,
while \ac{ECRA} with the \ac{RCB} reaches $T_{max-ECRA} = 1.01$ at
$G = 1.1\,\text{Erl}$. It is interesting to observe that the
throughput increase of \ac{ECRA} with respect to \ac{CRA} with the
\ac{SB} is 23\%, while it becomes 26\% when the \ac{RCB} is
considered, confirming the good performance of the proposed scheme
also in more practical situations.

\begin{figure}
\centering
\includegraphics[width=9.5cm]{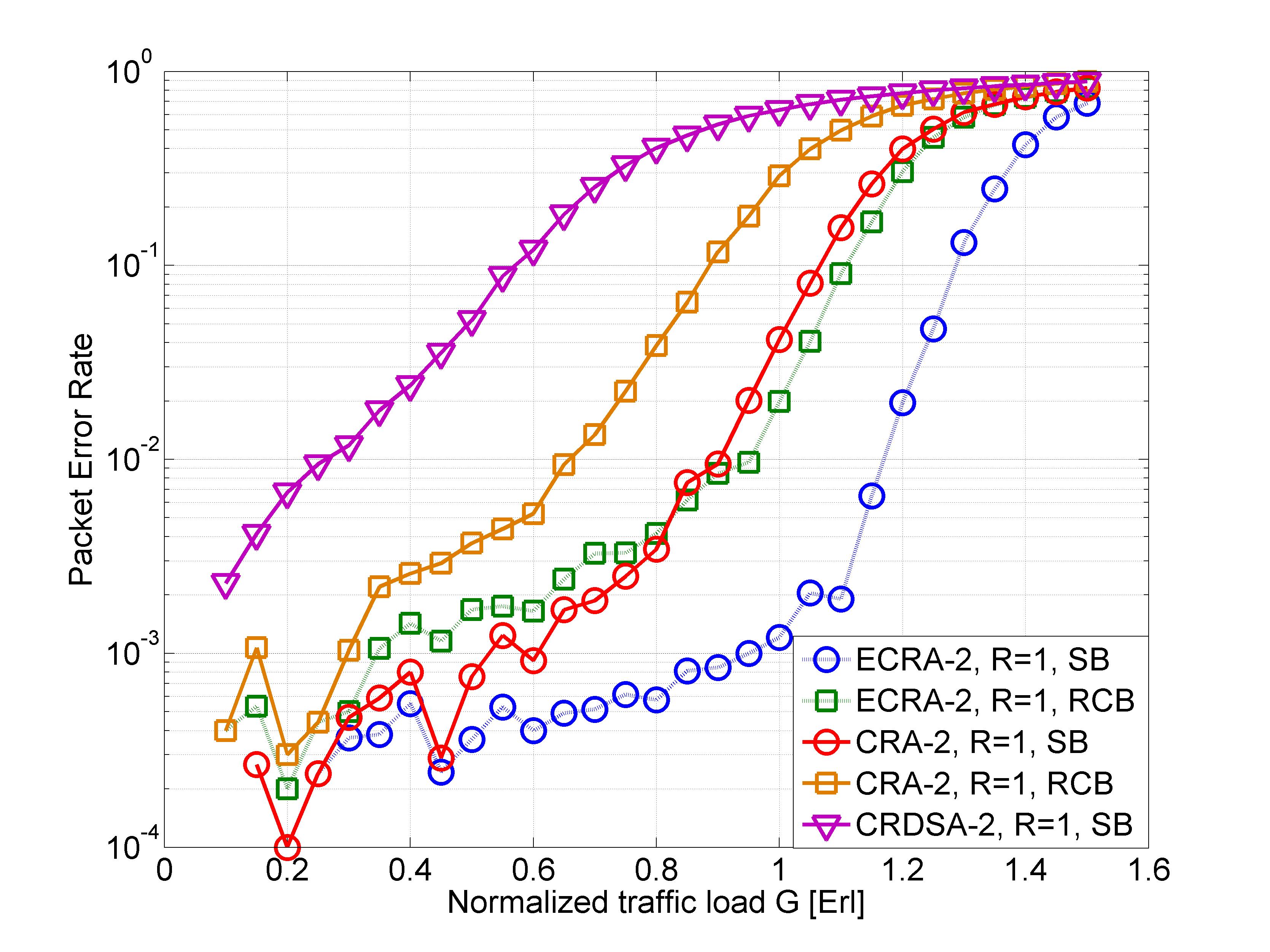}
\caption{\ac{CRDSA}, \ac{CRA} and \ac{ECRA} packet error rate comparison for rate $R=1$ and $SNR=10\,\text{dB}$}
\label{ECRA_R1_10_PER}
\end{figure}

In Fig. \ref{ECRA_R1_10_PER} the \ac{PER} behavior of the second set
of simulations is shown. We can observe that the minimum \ac{PER}
for all the considered simulations is similar and close to
$10^{-4}$.
This is due to the bound given by the number of simulated frames ($N_f=1000$).
When the \ac{SB} is considered, \ac{ECRA} can outperform \ac{CRA} by more than one order of
magnitude in the \ac{PER}, for $G \geq 0.9\,\text{Erl}$. The same
increase of performance can be found for the \ac{RCB}
simulations but for $G \geq 0.8\,\text{Erl}$.

\begin{figure}
\centering
\includegraphics[width=9.5cm]{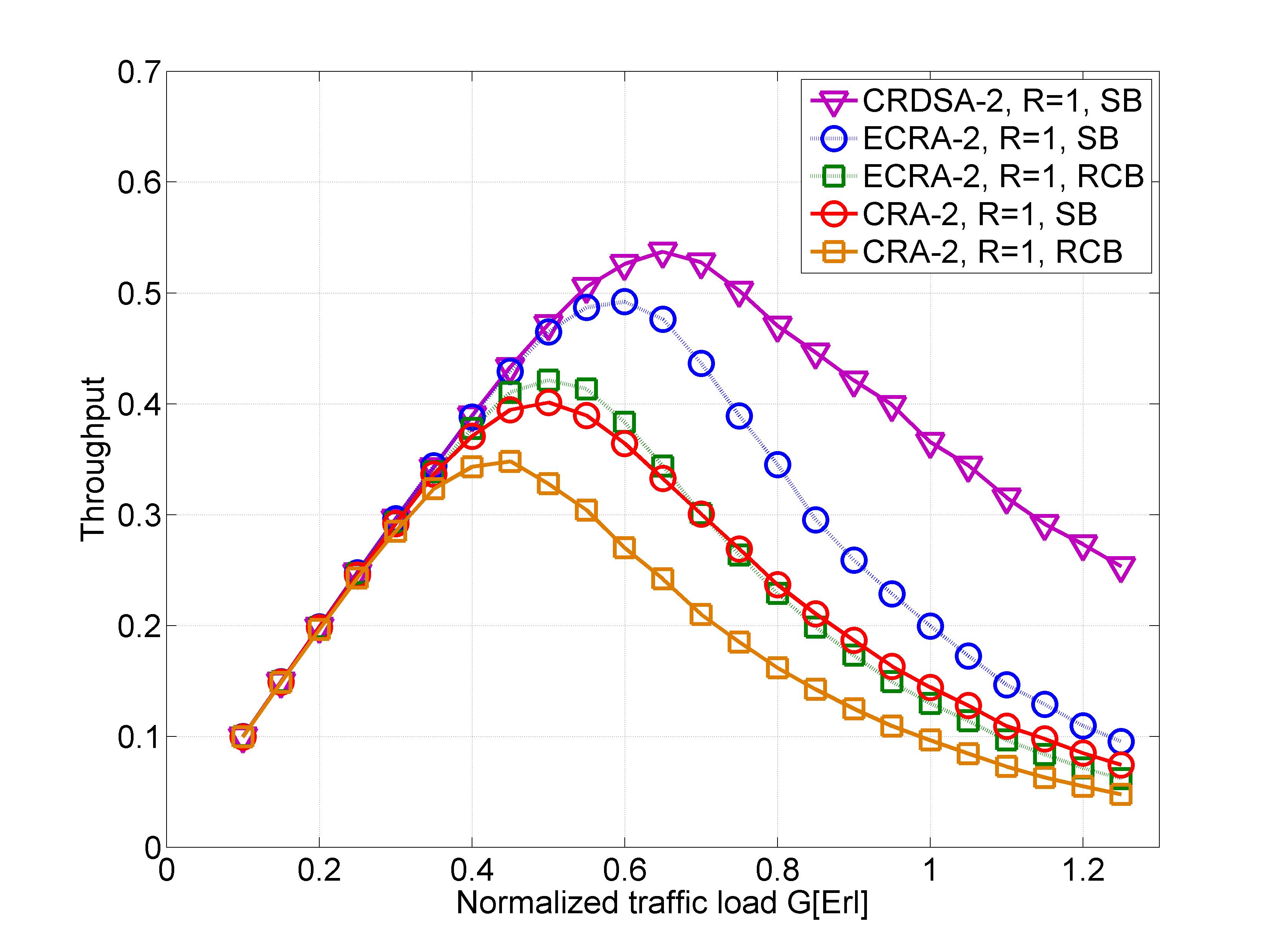}
\caption{\ac{CRDSA}, \ac{CRA} and \ac{ECRA} throughput comparison for rate $R=1$ and $SNR=2\,\text{dB}$}
\label{ECRA_R1_2_T}
\end{figure}

The third set of simulations provided are done for a rate $R=1$ and
with a reduced $SNR=2\,\text{dB}$, comparing the \ac{SB} with
the \ac{RCB}. In Fig. \ref{ECRA_R1_2_T} the throughput of
\ac{ECRA}, \ac{CRA} and \ac{CRDSA} are compared. It is \ac{CRDSA}
that reaches the best maximum throughput of $T_{max-CRDSA} \cong
0.53$ at $G=0.65\,\text{Erl}$. \ac{ECRA} with the \ac{SB}
reaches the maximum throughput $T_{max-ECRA} = 0.49$ at $G =
0.6\,\text{Erl}$ which is still close to \ac{CRDSA} and highly
outperforms \ac{CRA}. The throughput increment of \ac{ECRA} with
respect to \ac{CRA} with the \ac{SB} is 23\%, while it becomes
24\% when the \ac{RCB} is considered, in this second
case.

\section{Conclusions}

In this paper a novel \ac{RA} \ac{MAC} protocol has been presented.
Following the approach of \ac{CRA}, the \ac{ECRA} protocol exploits
the presence of multiple packet replicas, together with the nature
of occurring interference in Aloha-like channels combined with
strong channel coding and the \ac{SIC} process for resolving packet
collisions. Moreover it was shown how \ac{ECRA} attempts to resolve
most of the partial collisions among packets, with the creation of a
combined packet, generated from the lowest interfered parts of the
replicas sent within the frame. It has been shown mathematically
that this combined packet achieves always an equal or higher
\ac{SNIR} with respect to its corresponding replicas.

It has been also shown through numerical simulations that \ac{ECRA}
outperforms \ac{CRA} in all the considered scenarios for both the
throughput and the \ac{PER}. The simulations have further shown that
\ac{ECRA} can achieve up to 26\% of throughput gain compared to
\ac{CRA} when the \ac{RCB} is considered. Under the same conditions,
the \ac{PER} of \ac{ECRA} has a gain of one order of magnitude with
respect to the \ac{CRA} \ac{PER}.

\ifCLASSOPTIONcaptionsoff
  \newpage
\fi

\bibliographystyle{IEEEtran}
\bibliography{IEEEabrv,References}

\begin{thebibliography}{10}
\providecommand{\url}[1]{#1}
\csname url@samestyle\endcsname
\providecommand{\newblock}{\relax}
\providecommand{\bibinfo}[2]{#2}
\providecommand{\BIBentrySTDinterwordspacing}{\spaceskip=0pt\relax}
\providecommand{\BIBentryALTinterwordstretchfactor}{4}
\providecommand{\BIBentryALTinterwordspacing}{\spaceskip=\fontdimen2\font plus
\BIBentryALTinterwordstretchfactor\fontdimen3\font minus
  \fontdimen4\font\relax}
\providecommand{\BIBforeignlanguage}[2]{{%
\expandafter\ifx\csname l@#1\endcsname\relax
\typeout{** WARNING: IEEEtran.bst: No hyphenation pattern has been}%
\typeout{** loaded for the language `#1'. Using the pattern for}%
\typeout{** the default language instead.}%
\else
\language=\csname l@#1\endcsname
\fi
#2}}
\providecommand{\BIBdecl}{\relax}
\BIBdecl

\bibitem{Abramson1970}
N.~Abramson, ``The aloha system: Another alternative for computer
  communications,'' in \emph{Proceedings of the 1970 Fall Joint Comput. Conf.,
  AFIPS Conf.}, vol.~37, Montvale, N.~J., 1970, pp. 281--285.

\bibitem{Roberts1975}
L.~G. Roberts, ``Aloha packet system with and without slots and capture,''
  \emph{SIGCOMM Comput. Commun. Rev.}, vol.~5, pp. 28--42, April 1975.

\bibitem{Casini2007}
E.~Casini, R.~De~Gaudenzi, and O.~Herrero, ``Contention resolution diversity
  slotted {ALOHA} ({CRDSA}): An enhanced random access scheme for satellite
  access packet networks,'' \emph{Wireless Communications, IEEE Transactions
  on}, vol.~6, no.~4, pp. 1408 --1419, April 2007.

\bibitem{Kissling2011a}
C.~Kissling, ``Performance enhancements for asynchronous random access
  protocols over satellite,'' in \emph{Communications (ICC), 2011 IEEE
  International Conference on}, june 2011, pp. 1 --6.

\bibitem{Choudhury1983}
G.~Choudhury and S.~Rappaport, ``Diversity aloha--a random access scheme for
  satellite communications,'' \emph{Communications, IEEE Transactions on},
  vol.~31, no.~3, pp. 450 -- 457, March 1983.

\bibitem{Liva2011}
G.~Liva, ``Graph-based analysis and optimization of contention resolution
  diversity slotted aloha,'' \emph{Communications, IEEE Transactions on},
  vol.~59, no.~2, pp. 477 --487, february 2011.

\bibitem{RioHerrero2009}
O.~del Rio~Herrero and R.~D. Gaudenzi, ``A high-performance {MAC} protocol for
  consumer broadband satellite systems,'' \emph{IET Conference Publications},
  vol. 2009, no. CP552, pp. 512--512, 2009.

\bibitem{RioHerrero2008}
------, ``A high efficiency scheme for quasi-real-time satellite mobile
  messaging systems,'' in \emph{Signal Processing for Space Communications,
  2008. SPSC 2008. 10th International Workshop on}, 2008, pp. 1 --9.

\bibitem{ZigZag}
S.~Gollakota and D.~Katabi, ``Zigzag decoding: Combating hidden terminals in
  wireless networks,'' \emph{SIGCOMM'08}, pp. 159--170, 2008.

\bibitem{SigSag}
A.~G.~D. A.~S.~Tehrani and M.~J. Neely, ``Sigsag: Iterative detection through
  soft message-passing,'' \emph{Proceedings, IEEE INFOCOM}, 2011.

\bibitem{Gallager1968}
R.~G. Gallager, \emph{Information Theory and Reliable Communication}.\hskip 1em
  plus 0.5em minus 0.4em\relax Wiley, 1968.

\end{thebibliography}
\end{document}